# The use of the Cherenkov radiation and the fluorescence light to calibrate the energy of giant air showers[*]

L.G. Dedenko[1], D.A. Podgrudkov[1], T.M. Roganova[2], G.F. Fedorova[2], E.Yu. Fedunin[3], G.P. Shozieev[1]

[1] Faculty of physics, M.V. Lomonosov Moscow State University, Moscow 119992, Leninskie Gory, Russia
[2] D.V. Skobeltsin Institute of Nuclear Physics, M.V. Lomonosov Moscow State University, Moscow 119992, Leninskie Gory, Russia
[3] Research&Development Center REAGENT, Moscow 119991, Kosygin avenu, Russia

(Dated: March 1, 2007)

## Abstract

In terms of the quark-gluon string model the analysis of the classic procedure to estimate the energy of giant air showers with help of the parameter s(600) (a density of energy deposition in the scintillator at a distance of 600 m from the shower core) have been carried out. Simulations of the signal s(600) with help of the CORSIKA code in terms of the hybrid scheme show energy estimates which are approximately a factor of 1.6 times lower than adopted at the Yakutsk array. The energy estimates calculated with the help of the Cherenkov radiation coincide with the experimental data. Simulations of deposited energy distributions in the atmosphere with help of the GEANT4 code and the CORSIKA code show that more than 20% of this energy may be deposited at distances above 100 m from the shower axis.

PACS number: 9650.sd

INTRODUCTION

Investigations of the primary cosmic radiation (PCR) at ultra-high energies and particularly in the energy interval where the Greizen-Zatsepin-Kuzmin (GZK) effect [1,2] is important show so far no definite conclusion about sharp decreasing of the primary particle flux at energies above $10^{20}$ eV. Data of the Akeno Giant Air Shower Array (AGASA) show 11 events at energies above $10^{20}$ eV [3] if the observed dependence of the signal s(600) on the zenith angle θ is used that is well above the expected value due to the GZK effect. If calculated dependence of the signal s(600) on the zenith angle is used (though some questions may be put to this simulations) then the observed number decreases to 5-6 events with energies above $10^{20}$ eV [4], also above the expected value. Four events with such ultra-high energies have been observed at the Yakutsk array in accordance with data [3,4] because the acceptance area of the Yakutsk array is several times less than the area of the AGASA array [3,4]. At the same time the High Resolution (HiRes) collaboration have been claimed that they do see the cutoff of the energy spectra in accordance with the GZK effect [6]. Data of the new Pierre Auger Observatory (PAO) array which is under construction can neither support nor reject the GZK effect due to large systematic errors (~50%) [7]. The energy estimates used at various arrays are different. So they should be compared and analyzed to produce the same energy spectrum from the data. The classic approach to get an energy estimate with help of the parameter s(600) is used both at the Yakutsk and the AGASA arrays. This parameter shows the deposited energy inside the plastic scintillator at a distance of 600 m from the shower axis and is measured in units of density of vertical equivalent muons (VEM/m$^2$) which can produce the same energy deposition. The calorimetric approach have been used at the Yakutsk array: the parameter s(600) is calibrated with help of the Cherenkov radiation. Simulations show that both the density Q(400) of these radiation at a distance of 400 m from the shower axis and its total flux Φ are proportional to the energy $E_0$ of the primary particle. The calculated dependence of the energy $E_0$ on the parameter s(600) have been used at the AGASA array. The fluorescence light produced in the atmosphere have been exploited to measure the energy $E_0$ at the HiRes array and to calibrate the parameter s38 (a signal in a water tank at a distance of 1000 m from the shower axis from the inclined showers with the zenith angle of 38º) at the PAO array. The muon component of a shower may also be used to estimate the energy $E_0$. In this paper the energy estimates simulated on the base of various components of the giant air shower (GAS) are presented.

## 1. THE APPROACH OF THE PARAMETER S(600)

The standard formula to estimate the energy of the GAS obtained by the calorimetric approach at the Yakutsk array [5] show the dependence:

$$E_0 = (4.8 \pm 1.2) \times 10^{17} s(600)^{0.98}. \qquad (1)$$

Here the energy $E_0$ is expressed in eV and the signal $s(600)$ – in VEM/m$^2$. We have made some new efforts to simulate this dependence.

First, the responses of the scintillator detectors used at the Yakutsk array on the shower particles have been calculated with help of the code GEANT4 [8]. If any particle of a type k (1 – electron, 2 – gamma or 3 – muon) with an energy E hits a detector at a distance R from the shower axis and at the zenith angle θ then the response $f_k(E, \theta, R)$ depends on these variables. Then the development of the GAS in the atmosphere have been simulated with help of the code CORSIKA [9] with the parameter $\varepsilon = 10^{-6}$ of the "thinning" procedure [10]. This procedure sets an energy limit $E_l = \varepsilon E_0$. Only particles with energies above this limit are followed in detail. In every interaction if energy $E_i$ of secondary particles $E_i < E_l$ then only one particle with energy E is sampled and some weight $w = \sum E_i / E$ is assigned to this particle. The weight w is increasing for particles with low energy E. Some artificial fluctuations are introduced due to the large values of w.

Finally, the signal s(R) at a distance of R from the shower axis can be calculated evidently as a sum of responses of all shower particles:

$$s(R) = \sum_{k=1}^{3} \sum_{i=1}^{I_k} w_{ki} f_k(E_i, \theta_i, R), \qquad (2)$$

where $I_k$ is a number of the k-particles which hit a detector, $w_{ki}$ is a weight of a particle of the k type with energy $E_i$ and the zenith angle $\theta_i$. The formula (2) shows clearly how the artificial fluctuations are arisen. As at the level of observation particles have typically an energy E of the order of ~1 MeV, then $E_l \gg E$ and weights w are large that leads to considerable artificial fluctuations. To decrease these fluctuations we have elaborated the hybrid scheme of calculations [11]. First, in terms of this hybrid scheme the data base (DB) library have been simulated with help of the program EGS4 [14] which is included in the code CORSIKA. By the Monte Carlo method in the real atmosphere both the electron-induced and gamma-induced cascades with several energies E from the interval of 0.001-10 GeV and generated at various depths x (with the step of 50 g/cm$^2$) have been simulated and with help of the GEANT4 code the lateral distribution functions (LDF) $s_k(E, x, R)$ of signals from particles produced in of these cascades at the level of observation $x_0$ have been calculated. Then the code CORSIKA have been used to follow the shower particles with energies above the threshold energy $E_{th} = 10$ GeV and to calculate source functions: all electrons and gammas with energies $E_j \leq 10$ GeV, which have been produced in a shower at depths $x_j$ and with weights $w_{ki}$. Let us denote the total numbers of electrons and gammas produced in a shower as $J_1$ and $J_2$ accordingly. In this case the energy $E_j \leq 10$ GeV is considerably above the value of ~1 MeV and weights $w_{kj}$ are not so large. So the formula (2) should be replaced by the following expression:

$$s(R) = \sum_{k=1}^{2} \sum_{j=1}^{J_k} w_{kj} s_k(E_j, x_j, R) + \sum_{i}^{I_3} w_{3i} f_3(E_i, \theta_i, R) . \qquad (3)$$

The functions $s_k(E_j, x_j, R)$ are estimated with help of interpolations from the DB library. It is clear that muons are accounted for as in (2).

Calculations in terms of the hybrid scheme gives the dependence:

$$E_0 = 3 \times 10^{17} s(600)^{0.99} , \qquad (4)$$

If only the code CORSIKA used and our calculations of detector responses with help of the GEANT4 code (see (2)) then we get the formula:

$$E_0 = 2.7 \times 10^{17} s(600)^{0.99} . \qquad (5)$$

The difference of estimates (4) and (5) by ~10% shows an accuracy of simulations. We believe that the estimate (4) is more accurate because not so high weights w are used in (3) in comparison with (2). This estimate (4) is by a factor of 1.6 times less than (1). It was shown [13] that the energy spectra of the primary cosmic rays observed both at the HiRes and the Yakutsk installations would be in accordance if this estimate (4) is used to interpret the Yakutsk data. But this fact does not mean that the estimate (4) calculated in terms of the model QGSJET01 [14] is more accurate than the experimental estimate (1) which have been calibrated with help of the Cherenkov radiation because the model used may not describe well the real physical processes at ultra-high energies. To agree the energy spectra [5] and [6] it was suggested [15,16] to decrease the energy estimate used in [5] and to increase the estimate used in [6] by ~30%. As it was shown there are some reasons to decrease the estimate (1) used in [5]. But to increase the estimate used in [6] there are not many possibilities. Some analysis of the energy estimate with help of the fluorescence light is given in the 3-rd point of our paper.

## 2. THE CALIBRATION OF ENERGY ESTIMATES WITH HELP OF THE CHERENKOV RADIATION

Also in terms of the hybrid scheme first the DB library have been calculated for the Cherenkov radiation [17] as in the case of simulations of the signal s(600). To create this DB library the lateral distribution functions $Q_k(E, x, R)$ of the Cherenkov radiation for distances of 5–1000 m from the shower axis have been simulated in cascades induced in the atmosphere by electrons and gammas with several energies E from the interval of 0.02–10 GeV and generated at various depths x. The calculations have been carried out for the Cherenkov radiation with the wave lengths of 300–750 nm. It was assumed that the Cherenkov photons are sampled in the detector within the zenith angle θ=55°. To simulate the LDF of the Cherenkov radiation in the GAS the source functions of electrons and gammas (see above) estimated with help of the CORSIKA code have been used and calculations have been carried out with help of a formula like (3). In Fig.1 solid curves show the LDF of the Cherenkov radiation in GAS with energies $10^{17}$–$10^{20}$ eV (curves 1, 2, 3 and 4 accordingly). The dotted curves and points are some approximations used at the Yakutsk array and the data [18,19]. A good agreement of calculated results with the data is seen for the most important distances above 200 m from the shower axis. Some difference at smaller distances may be induced by uncertainties in the location of the shower axis. In Fig.2 the dependence of the parameter Q(400) – the density of the Cherenkov photons at a distance of 400 m from the shower axis – divided on the energy $E_0$ (in GeV) on the energy $E_0$ is shown. Points illustrate calculated results. Solid curve is some approximation of these results. The dotted curve shows the data [18]. It is seen that the calculated results agree with the data within some errors of 10–15%. It is evidently that the parameter $Q(400)$ may be used to estimate the energy $E_0$ of a shower. This parameter can also be used to calibrate the signals s(600) and the densities $\rho_\mu(600)$ of muons with energies above 1 GeV detected at a distance of 600 m from the shower axis. Fig.3 illustrates dependences of the parameter $Q(400)$ on the signals s(600) and on the muon densities $\rho_\mu(600)$ (solid lines) and on the data [19] (dotted lines). It can be seen that the calculated value of the signal s(600) is by a factor of ~1.6 times larger than the value of the measured signal at the same value of the parameter $Q(400)$ which is proportional to the energy of a shower. This agrees with the results shown above. It is interesting to note that calculated muon densities are in accordance with the data. It is important for the energy estimation of the very inclined showers. The calculated fractions of the energy deposited in the atmosphere are equal to 0.832, 0.787 and 0.747 for the vertical showers with energies $10^{18}$, $10^{19}$ and $10^{20}$ eV accordingly and agree with the data [18,19]. Evidently, the new analysis of the absolute calibration of detectors of the Cherenkov radiation is needed.

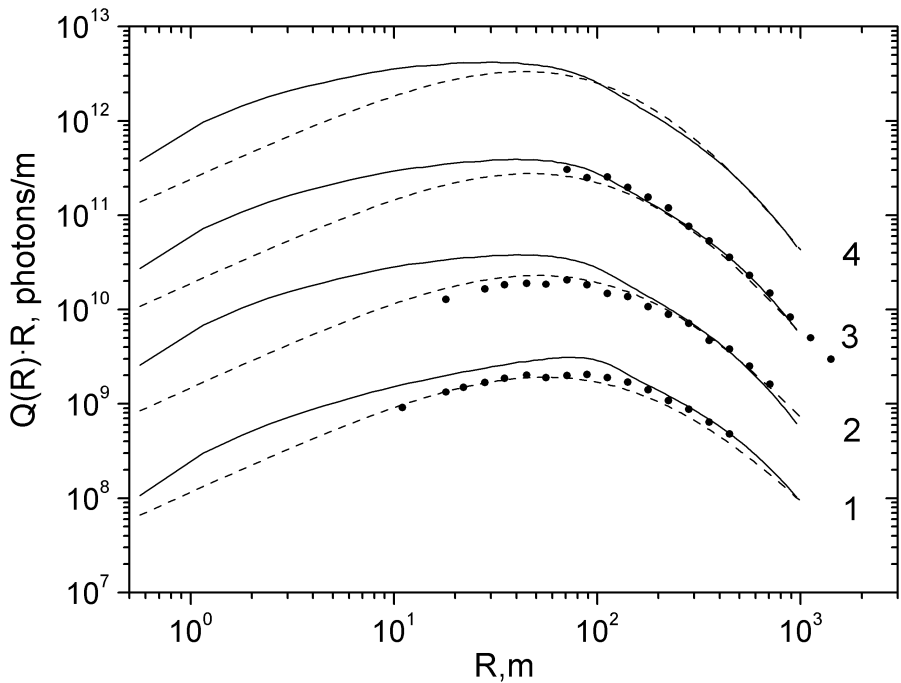

Fig.1. The lateral distribution functions of the Cherenkov radiation. Solid curves – calculations, dotted curves and points – the data [18,19]. 1 – $10^{17}$ eV, 2 – $10^{18}$ eV, 3 – $10^{19}$ eV, 4 – $10^{20}$ eV.

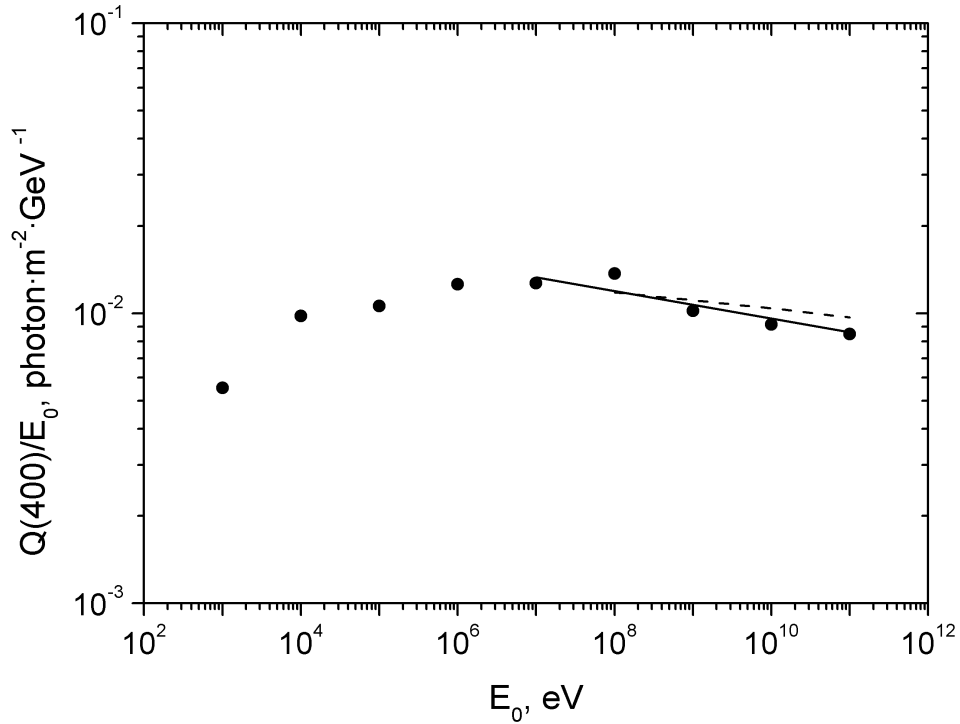

Fig.2. The dependence of the ratio $Q(400)/E_0$ on the energy $E_0$. Solid curve and points – calculations, dotted curve – the data [18].

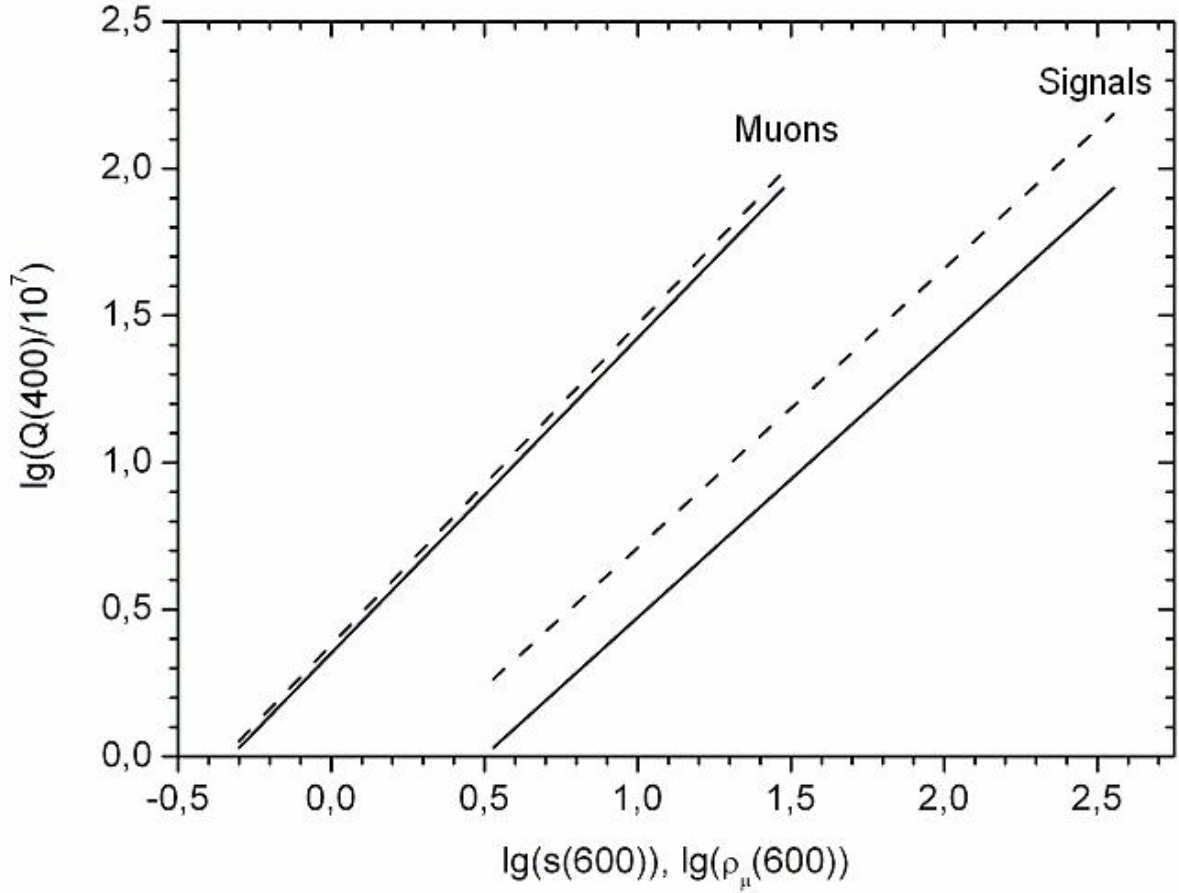

Fig.3. The dependence of the parameter Q(400) on the signal s(600) and the density $\rho_\mu(600)$ of muons with energies above 1 GeV. Solid curves – calculations, dotted lines – the data [18,19].

3. ENERGY ESTIMATES OF GAS WITH THE USE OF THE FLUORESCENCE LIGHT

First, the DB library for the fluorescence light have been simulated with the help of the GEANT4 code [20,21] contrary to the case of calculations of the signal s(600) and the Cherenkov radiation with the use of the CORSIKA code. To create this DB library the electron-photon cascades induced in the atmosphere by electrons and gammas with several energies E from the interval of 0.001 – 10 GeV at various depths x have been simulated by the Monte Carlo method and the densities of the deposited energy $\Delta E_k(E, x, R_{n-1}, R_n)$ inside the fixed volumes have been calculated. Then the lateral distributions of the energy $\Delta E$ deposited inside cylinders with the width of $R_n - R_{n-1} = 20$ m for the GAS have been calculated with the use of the formula like (3) and the source functions mentioned above. Fig.4 shows how the calculated values of this deposited energy ΔE depend on the distance R from the shower axis for showers with energies of $10^{18}$, $10^{19}$ and $10^{20}$ eV (curves 1, 2 and 3 accordingly). As it is seen from this figure the lateral distributions of the deposited energy are rather wide. It is convenient to characterize this width by the value of the radius R of the cylinder inside which the fraction α of the total energy of a shower have been deposited. Fig.5 shows how this fraction α of the total energy deposited inside the cylinder with the radius R depends on this radius for showers with energies of $10^{18}$, $10^{19}$ and $10^{20}$ eV (curves 1, 2 and 3 accordingly).

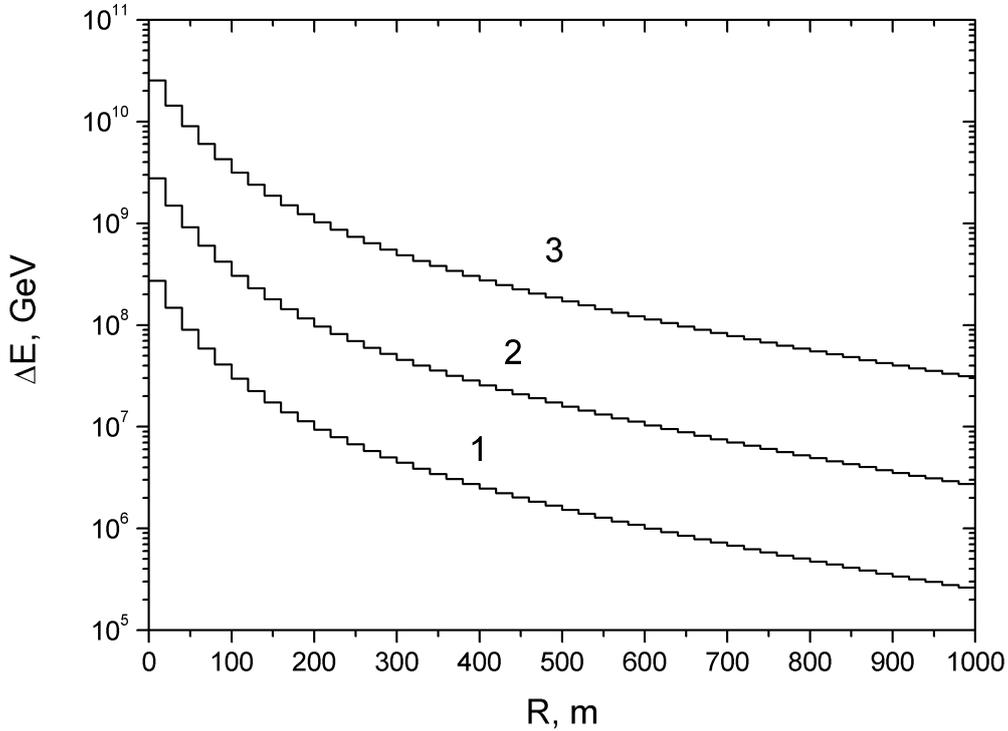

Fig.4. The lateral distributions of the deposited energy for various energies of showers. 1 – $10^{18}$ eV, 2 – $10^{19}$ eV, 3 – $10^{20}$ eV.

It is seen from this figure that the fraction of 90% of the deposited energy are confined in the cylinders with radii of 210, 250 and 290 m for curves 1, 2 and 3 accordingly. Let us note that the fraction of 95% of the deposited energy is confined in the cylinder with the radius of ~500 m. It should be noted that only ~80% of the fluorescence light which is proportional to the deposited energy is detected if the axis location do not exceed a distance of ~5 km for the sampling angle of ~1.3°. The detected light exceeds 95% of its total value for the same sampling angle only for distances above 25 km. The calculations [22] show that only 10% of the energy is deposited at distances above 100 m from the shower axis. In contrast, our simulations show that this value should be above ~20%. The main advantages of our estimates are the use of the hybrid scheme with relatively small values of weights w and the use of the code GEANT4 with the threshold energies of secondary particles which are practically equal to zero. In case of the CORSIKA code used in [22] the threshold energy $E_{min}$ is equal to 0.1 MeV. It is the low energy particles (mainly gammas) which spread to large distances from the shower axis. This is the reason why the lateral distribution functions of the deposited energy calculated in [22] are more steep than shown at Fig.4. Let us note that some fluorescence light may be not detected due to the fixed signal - to - noise ratio. Besides, our calculations show that nearly 20% of the total energy of a shower is not detected in the vertical showers because this energy are transferred beyond the level of observation. Thus, some underestimation of the shower energy $E_0$ is possible in experiments which use the fluorescence light. To get the detailed estimates some new simulations are needed with help of the GEANT4 code e.g. for the HiRes array.

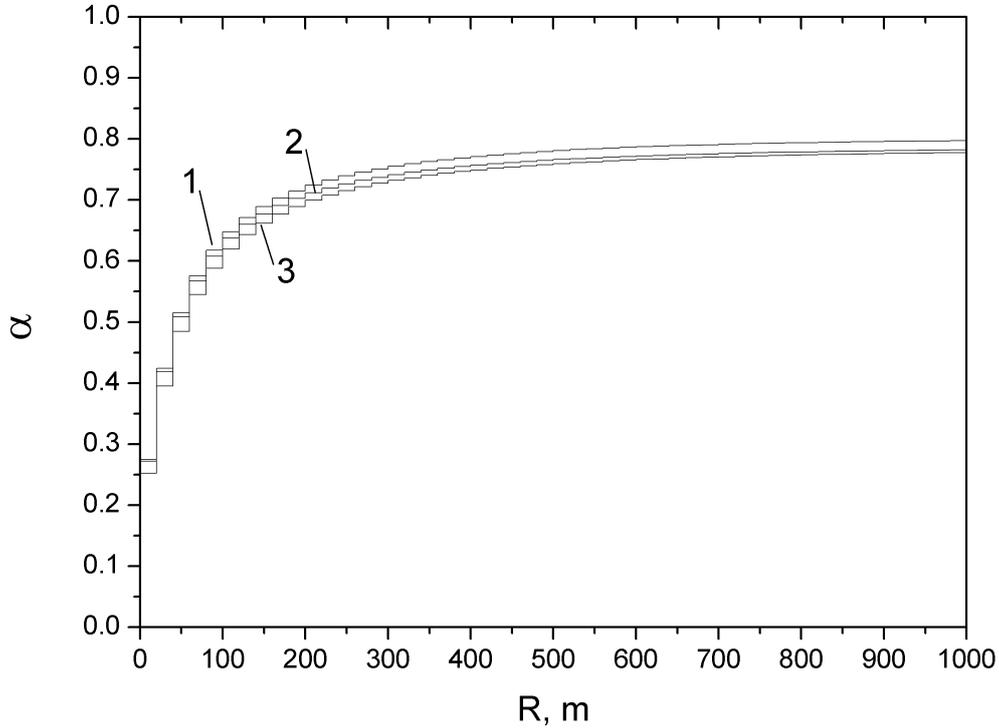

Fig.5. The dependence of the fraction α of the deposited energy on the distance R from the shower axis for showers with various energies. $1 - 10^{18}$ eV, $2 - 10^{19}$ eV, $3 - 10^{20}$ eV.

CONCLUSION

Our calculations in terms of the QGSJET01 model with help of the GEANT4 code, the EGS4 code and the CORSIKA code show that the signal s(600) may be nearly 1.6 times less that adopted at the Yakutsk experiment [5] at the same energy of the vertical shower induced by the primary protons. If these calculated values of signals are used then the intensity of the primary cosmic rays observed both at the Yakutsk array and the HiRes are agreed [5]. Simulations of the LDF and the parameter Q(400) of the Cherenkov radiation are in accordance with the Yakutsk data. The calculated and experimental calibrations of the signal s(600) with the use of the parameter Q(400) are considerably (by 1.6 times) disagreed. At the same time both the calculated and experimental densities of muons with energies above 1 GeV observed at a distance of 600 m from the shower axis are in accordance at the same value of the parameter Q(400). It is important for the correct estimation of the energy of the inclined showers. Our simulations with the use of the GEANT4 code show that the fluorescence light is produced up to distances above ~500 m from the shower axis.

Acknowledgements.
Authors thank G.T. Zatsepin and V.A. Kuzmin for valuable comments and the INTAS (grant 03-51-5112) and the LSS (grant 5573.2006.2) for the financial support.